\documentclass[floatfix, preprint, showpacs, showkeys, preprintnumbers, nofootinbib, superscriptaddress]{revtex4-1}
\usepackage[utf8]{inputenc}
\usepackage[sort&compress]{natbib}
\usepackage{ulem}
\usepackage{bm}
\usepackage{overpic}
\usepackage{times}
\usepackage{amssymb,amsbsy,amsmath,amsfonts}
\usepackage{graphicx}
\usepackage{float}
\usepackage{color}
\usepackage{morefloats}
\usepackage{rotating}
\usepackage{srcltx}
\usepackage{slashed}
\usepackage{subfigure}
\usepackage{multirow}
\usepackage{verbatim}
\usepackage{hyperref}
\usepackage{tabularx}

\begin{document}

\title{Understanding $P_{cs}(4459)$ as a hadronic molecule in the $\Xi_b^-\to J/\psi \Lambda K^-$ decay}

\author{Jun-Xu Lu}
\affiliation{School of Space and Environment,  Beihang University, Beijing 102206, China}
\affiliation{School of
Physics,  Beihang University, Beijing 102206, China}

\author{Ming-Zhu Liu}
\affiliation{School of Space and Environment,  Beihang University, Beijing 102206, China}
\affiliation{School of
Physics,  Beihang University, Beijing 102206, China}

\author{Rui-Xiang Shi}
\affiliation{School of
Physics,  Beihang University, Beijing 102206, China}

\author{Li-Sheng Geng}
\email[Corresponding author: ]{lisheng.geng@buaa.edu.cn}
\affiliation{School of
Physics,  Beihang University, Beijing 102206, China}
\affiliation{Beijing Key Laboratory of Advanced Nuclear Materials and Physics, Beihang University, Beijing 102206, China }
\affiliation{School of Physics and Microelectronics, Zhengzhou University, Zhengzhou, Henan 450001, China }

\begin{abstract}
Recently, the LHCb Collaboration reported on the evidence for a hidden charm pentaquark state with strangeness, i.e., $P_{cs}(4459)$, in the $J/\psi\Lambda$ invariant mass distribution of the $\Xi_b^-\to J/\psi \Lambda K^-$ decay. In this work, assuming that $P_{cs}(4459)$ is a $\bar{D}^*\Xi_c$ molecular state, we study this decay via triangle diagrams
$\Xi_b\rightarrow \bar{D}_s^{(*)}\Xi_c\to (\bar{D}^{(*)}\bar{K})\Xi_c\to P_{cs} \bar{K}\to (J/\psi\Lambda) \bar{K}$. Our study shows that the production yield of a spin 3/2 $\bar{D}^*\Xi_c$ state is approximately one order of magnitude larger than that of a spin $1/2$ state due to the interference of $\bar{D}_s\Xi_c$ and $\bar{D}_s^*\Xi_c$ intermediate states. We obtain
a model independent constraint on the product of couplings $g_{P_{cs}\bar{D}^*\Xi_c}$ and $g_{P_{cs}J/\psi\Lambda}$. With the predictions of two particular molecular models as inputs, we calculate the branching ratio of $\Xi_b^-\to (P_{cs}\to)J/\psi\Lambda K^- $ and compare it with the experimental measurement. We further predict the lineshape of this decay which could be useful to future experimental studies.
\end{abstract}


\maketitle

\section{Introduction}
In 2015, two pentaquark states $P_c(4380)$ and $P_c(4450)$ were observed in the $J/\psi p$ invariant mass distribution of the $\Lambda_b \rightarrow J/\psi p K^-$ decay by the LHCb Collaboration~\cite{Aaij:2015tga}, which have long  been anticipated theoretically~\cite{Wu:2010jy,Wu:2010vk,Wang:2011rga,Yang:2011wz,Yuan:2012wz,Wu:2012md,Garcia-Recio:2013gaa,Xiao:2013yca,Uchino:2015uha,Karliner:2015ina}. Since then, a large amount of theoretical works have been performed to understand their nature. The most popular interpretations include $\bar{D}^{(*)}\Sigma_c^{(*)}$ molecular states~\cite{Roca:2015dva,He:2015cea,Xiao:2015fia,Chen:2015loa,Chen:2015moa,Burns:2015dwa,Shen:2016tzq,Geng:2017hxc,Liu:2018zzu}, compact pentaquark states~\cite{Maiani:2015vwa,Lebed:2015tna,Wang:2015epa}, and kinematical effects~\cite{Guo:2015umn,Liu:2015fea}.
The experimental results were updated in 2019 with a data sample of almost ten times larger~\cite{Aaij:2019vzc}. A new narrow state, $P_c(4312)$ was discovered. More interestingly, the original $P_c(4450)$ state  splits into two states, $P_c(4440)$ and $P_c(4457)$. The masses and widths of these states are tabulated in Table~\ref{Tab:Para}. After the 2019 update,
 the pentaquark states look more like  $\bar{D}^{(*)}\Sigma_c^{(*)}$ molecules~\cite{Chen:2019bip,Chen:2019asm,He:2019ify,Liu:2019tjn,Shimizu:2019ptd,Xiao:2019mvs,Guo:2019kdc,Fernandez-Ramirez:2019koa,Xiao:2019aya,Yamaguchi:2019seo,Wu:2019rog,Valderrama:2019chc,Liu:2019zvb,Lin:2019qiv}, but again non-molecular interpretations are possible, such as compact pentaquark states~\cite{Ali:2019clg,Wang:2019got,Ali:2019npk}  and even double triangle singularities~\cite{Nakamura:2021qvy}.

Most recently, the LHCb Collaboration reported on the first evidence for a structure in the $J/\psi\Lambda$ invariant mass distribution of the $\Xi_b^- \rightarrow J/\psi\Lambda K^-$ decay~\cite{Aaij:2020gdg}, hinting at the existence of a pentaquark state with strangeness, i.e., $P_{cs}(4459)$.  It should be noted that the existence of pentaquark states with strangeness was predicted together with their non-strange counterparts  in the molecular picture~\cite{Wu:2010jy,Wu:2010vk}. The $P_{cs}(4459)$ state is located  close to the $\bar{D}^*\Xi_c$ threshold, leading naturally to a molecular interpretation~\cite{Peng:2020hql,Xiao:2021rgp,Zhu:2021lhd,Dong:2021juy,Liu:2020hcv,Chen:2021tip,Liu:2020hcv,Xiao:2021rgp}. One interesting point to be noted is that in addition to the four $P_c$'s discovered experimentally, there may be three more candidates which strongly couple to $\bar{D}^*\Sigma_c^*$ with $J^P=\frac{1}{2}^-, \frac{3}{2}^-, \frac{5}{2}^-$, as dictated by the heavy quark spin symmetry (HQSS)~\cite{Liu:2019tjn,Xiao:2019aya,Yamaguchi:2019seo,Liu:2019zvb}.  As for pentaquark states with strangeness, one expects 10 of them~\cite{Xiao:2019gjd,Wang:2019nvm,Liu:2020hcv,Xiao:2021rgp}.

 \begin{table}[htpb]
 \centering
 \caption{Resonance parameters of the newly discovered pentaquark states and their production ratios, defined as $R=\frac{\mathcal{B}(\Lambda_b(\Xi_b)\rightarrow P_c(P_{cs})\bar{K})\mathcal{B}(P_c(P_{cs})\rightarrow J/\psi p(\Lambda))}{\mathcal{B}(\Lambda_b(\Xi_b)\rightarrow \bar{K}J/\psi p(\Lambda))}$. \label{Tab:Para} }
 \begin{tabular}{ccccc}
 \hline\hline
 State & Mass (MeV) & Width (MeV) & \multicolumn{2}{c}{R($\%$)} \\
 \hline
 $P_c(4312)$ & $4311.9 \pm 0.7^{+6.8}_{-0.6}$  & $9.8 \pm 2.7 ^{+3.7}_{-4.5}$ & \multicolumn{2}{c}{$0.30 \pm 0.07^{+0.34}_{-0.09}$\cite{Aaij:2019vzc}}   \\
 $P_c(4380)$ &$4380\pm8\pm29$ & $205\pm18\pm86$ & \multicolumn{2}{c}{$ 8.4 \pm 0.7 \pm 4.2$\cite{Aaij:2015tga} } \\
 $P_c(4440)$ &$4440.3 \pm 1.3^{+4.1}_{-4.7}$  & $20.6 \pm 4.9 ^{+8.7}_{-10.1}$ & $1.11 \pm 0.33^{+0.22}_{-0.10}$\cite{Aaij:2019vzc} & \multirow{2}{*}{$4.1 \pm 0.5 \pm 1.1$\cite{Aaij:2015tga}}\\
 $P_c(4457)$ & $4457.3 \pm 0.6^{+4.1}_{-1.7}$  & $6.4 \pm 2.0 ^{+5.7}_{-1.9}$ & $0.53 \pm 0.16^{+0.15}_{-0.13}$\cite{Aaij:2019vzc}  & \\
 $P_{cs}(4459)$ & $4458.8\pm2.9^{+4.7}_{-1.1}$ & $17.3\pm6.5^{+8.0}_{-5.7}$ & \multicolumn{2}{c}{$2.7^{+1.9+0.7}_{-0.6-1.3}$\cite{Aaij:2020gdg}} \\
\hline\hline
 \end{tabular}
 \end{table}

In addition to the masses and widths of the pentaquark states, the LHCb Collaboration also reported the production yields of $P_c(4312)$, $P_c(4380)$, $P_c(4440)$, $P_c(4457)$, and $P_{cs}(4459)$, which are collected in Table~\ref{Tab:Para}.
One notes that the production yield for $P_c(4380)$ is one order of magnitude larger than that of $P_c(4312)$, which provides an explanation why  $P_c(4312)$ was not observed in 2015. However, we note that the sum of the production yields of $P_c(4440)$ and $P_c(4457)$ is only half of that of $P_c(4450)$, which may indicate that something is missing, maybe a new resonance as suggested in several  works~\cite{Geng:2017hxc,Burns:2019iih,Peng:2020gwk,Xu:2020gjl,Xu:2020gjl}. Clearly, understanding the production yields, particularly, the pattern shown in Table~\ref{Tab:Para}, will greatly improve our understanding of the pentaquark states.

In the present work,  we study the branching ratio of $\Xi_b^-\rightarrow P_{cs}K^-$.~\footnote{In Ref.~\cite{Ling:2021qzl}, a similar mechanism has been applied to study the  $D_s^{+}\to \pi^{+}\pi^0 \eta$  decay and it is shown that both the branching ratio and the $\pi^{+(0)}\eta$ lineshape  are well described. In particular, the large branching ratio of $D_s^+\to a_0^{+}\pi^{0}(a_0^{0}\pi^{+})$ is naturally explained, while for a pure $W$-annihilation process one would expect a much smaller value.} The present work differs from those of Refs.~\cite{Chen:2015sxa,Wu:2021dmq} in two ways. First, the weak production formalism is different from that of Ref.~\cite{Chen:2015sxa} (see also Ref.~\cite{Lu:2016roh}), which allows for a prediction of the absolute branching ratio of the $\Xi_b^-\to P_{cs} K^-\to J/\psi \Lambda K^-$ decay, within a molecular model. Compared to Ref.~\cite{Wu:2021dmq}, we use different parameterizations of form factors and predict the lineshape of the $\Xi_b^-\to P_{cs} K^-\to J/\psi \Lambda K^-$ decay.
 Taking two molecular models for the $P_{cs}(4459)$ state~\cite{Xiao:2021rgp}, we compare the so-obtained branching ratios with the experimental data.

The present work is organized as follows. In Sec.~II, we explain in detail the mechanism for the $P_{cs}(4459)$ production in the $\Xi_b^-$ decay, which involves a weak interaction part and a strong interaction part. In Sec.~III, we present the numerical results and compare with the experimental data, followed by a short summary in Sec.~IV.

\section{Theoretical Framework}

\begin{figure*}[htpb]
\centering
\begin{tabular}{cc}
{\includegraphics[width=0.45\textwidth]{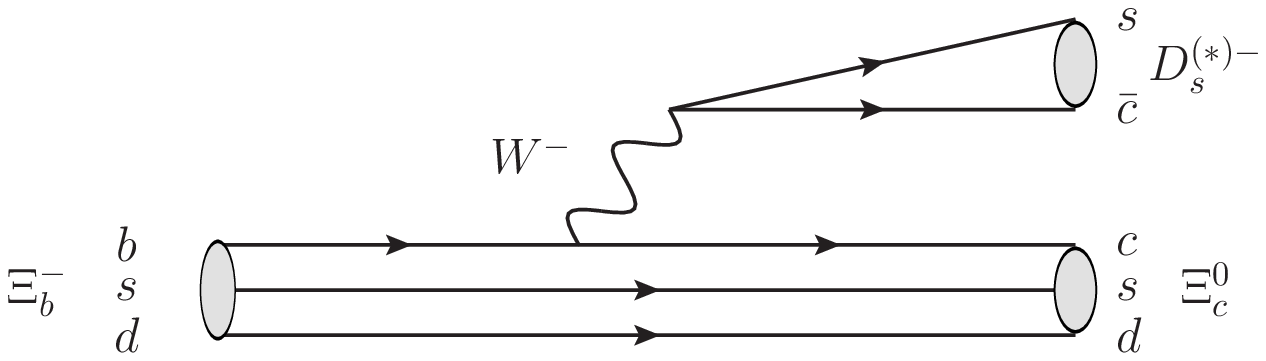}}&{\includegraphics[width=0.45\textwidth]{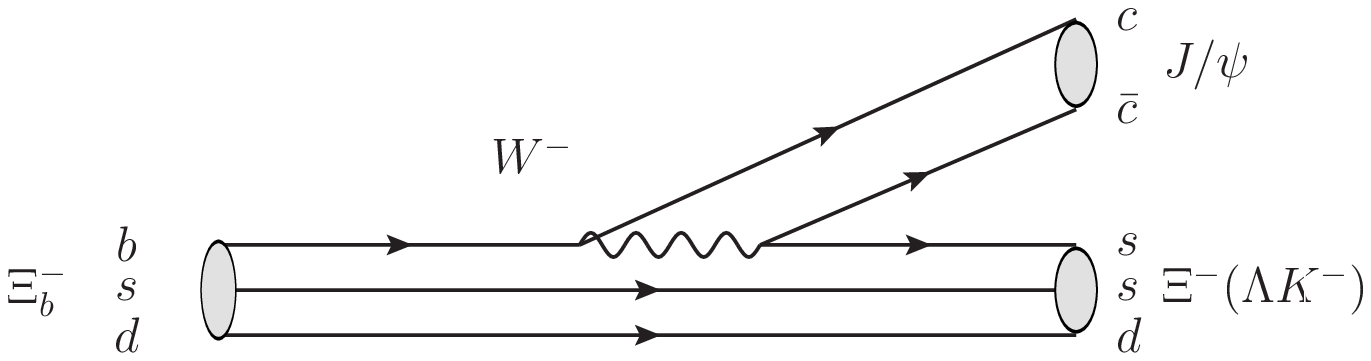}}\\
(a) & (b) \\
\end{tabular}
\caption{External W-emission (a) and internal W-conversion (b) mechanism for the $\Xi_b$ decay. }
\label{quarklevel}
\end{figure*}

\begin{figure}[h!]
\begin{center}
\begin{tabular}{cc}
\begin{minipage}[t]{0.45\linewidth}
\begin{center}
\begin{overpic}[scale=.6]{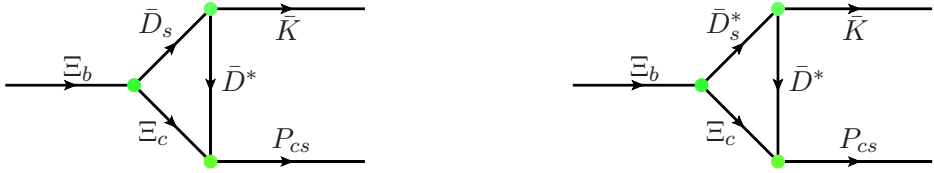}
		\put(74,6){$P_{cs}$}
		
		\put(37,8){$\Xi_{c}$}
		
		\put(37,38){$\bar{D}_{s}$}
		
		\put(16,26){$\Xi_{b}$ }
		\put(75,36){$\bar{K}$} \put(60,22){$\bar{D}^{\ast}$}
\end{overpic}
\end{center}
\end{minipage}
&
\begin{minipage}[t]{0.45\linewidth}
\begin{center}
\begin{overpic}[scale=0.6]{triangle.eps}
		\put(74,6){$P_{cs}$}
		
		\put(37,8){$\Xi_{c}$}
		
		\put(37,38){$\bar{D}_{s}^{\ast}$}
		
		\put(16,26){$\Xi_{b}$ }
		\put(75,36){$\bar{K}$} \put(60,22){$\bar{D}^{\ast}$}
\end{overpic}
\end{center}
\end{minipage}
\end{tabular}
\caption{Triangle diagrams for the $\Xi_b^-\to P_{cs}K^-$ decay.}
\label{decaywidth}
\end{center}
\end{figure}

Assuming that  $P_{cs}(4459)$ is a molecule mainly composed of $\bar{D}^*\Xi_c$, the $\Xi^-_b\rightarrow P_{cs} K^-$ decay can proceed as shown in diagram (a) of Fig.~\ref{quarklevel}. The $\Xi_b$ state first decays into $\Xi_c$ by emitting a $W^-$ boson which is then converted into  a pair of $\bar{c}s$, which after hadronization turns into a $D_s^{(*)}$. Next the $D_s^{(*)}$ meson emits a kaon and a $\bar{D}^*$.  The final state interaction of $\bar{D}^*\Xi_c$ dynamically generates the $P_{cs}(4459)$ state which then decays into $J/\psi\Lambda$, as shown in Fig.~\ref{decaywidth}.

In addition to the $W$-emission diagram discussed above, the $\Xi_b^-$ decay can also proceed via the internal $W$-exchange mechanism shown in diagram (b) of Fig.~\ref{quarklevel}. The $ssd$ cluster can either directly hadronize into a $\Xi^-$ or, by picking up a pair of $q\bar{q}$ from the vacuum,  hadronizes into  $\Lambda K^-$. The former is indeed a main decay channel of $\Xi_b^-$~\cite{Zyla:2020zbs}, while the latter has been studied in  Ref.~\cite{Chen:2015sxa}.

\subsection{Branching ratio of $\Xi_b\to P_{cs}K^-$}
In the following, we describe how to calculate the diagrams of Fig.~\ref{decaywidth}. The effective Lagrangian responsible for the $\Xi_b\rightarrow\Xi_c \bar{D}_s^{(*)}$ decay reads
\begin{equation}\label{Eq:weakdecayV1}
\begin{split}
    \mathcal{L}_{\Xi_b\Xi_cD_s}&=i\bar{\Xi}_c(A+B \gamma_5)\Xi_bD_s,\\
    \mathcal{L}_{\Xi_b\Xi_cD_s^*}&=\bar{\Xi}_c(A_1\gamma_\mu\gamma_5+A_2\frac{p_{2\mu}}{m}\gamma_5+B_1\gamma_\mu +B_2\frac{p_{2\mu}}{m})\Xi_bD_s^{*\mu}.
\end{split}
\end{equation}
The $A_1$, $A_2$, $B_1$, $B_2$, $A$, and $B$ can be expressed with the six form factors describing the $\Xi_b\rightarrow\Xi_c$ transition~\cite{Cheng:1996cs} as~\footnote{Here we adopt the convention for the form factors of Ref.~\cite{Faustov:2018ahb} in which there exists an extra minus in front of $f_2^A$ and $f_2^V$.}
\begin{equation}\label{Eq:weakdecayV2}
\begin{split}
    A&=\lambda f_{D_s}[(m-m_2)f_1^V+\frac{m_1^2}{m}f_3^V], \quad
    B=\lambda f_{D_s}[(m+m_2)f_1^A-\frac{m_1^2}{m}f_3^A], \\
    A_1&=-\lambda f_{D_s^*}m_1[f_1^A-f_2^A\frac{m-m_2}{m}], \quad\quad
    B_1=\lambda f_{D_s^*}m_1[f_1^V+f_2^V\frac{m+m_2}{m}], \\
    A_2&=2\lambda f_{D_s^*}m_1f_2^A, \quad\quad\quad\quad\quad\quad\quad\quad
    B_2=-2\lambda f_{D_s^*}m_1f_2^V,
\end{split}
\end{equation}
where $\lambda=\frac{G_F}{\sqrt{2}}V_{cb}V_{cs}a_1$ with $a_1=1.07$~\cite{Li:2012cfa}. The decay constants $f_{D_s^{(*)}}$ for $\bar{D}_s$ and $\bar{D}_s^*$ are set to be 0.247 GeV and $m, m_1, m_2$ refer to the masses of $\Xi_b$, $\bar{D}_s^{(*)}$, and $\Xi_c$ respectively.

Following the double-pole parametrization proposed in Ref.~\cite{Gutsche:2015mxa}, one can rewrite the form factors as
\begin{equation}
   f_i^{V/A}(q^2)=F_i^{V/A}(0)\frac{\Lambda_1^2}{q^2-\Lambda_1^2}\frac{\Lambda_2^2}{q^2-\Lambda_2^2}.
\end{equation}
Fitting to the results of the relativistic quark-diquark model~\cite{Faustov:2018ahb}, we can obtain the values of $F(0)$, $\Lambda_1$, and $\Lambda_2$, which are tabulated in Table~\ref{Tab:FormFactor}.
 \begin{table}[ttt]
 \centering
 \caption{Parameters $F(0)$, $\Lambda_1$, $\Lambda_2$ in the form factors of the $\Xi_b\rightarrow\Xi_c$ transition form factors. \label{Tab:FormFactor} }
 \begin{tabular}{ccccccc}
 \hline\hline
   & $F_1^V$ & $F_2^V$ & $F_3^V$ & $F_1^A$ &  $F_2^A$ & $F_3^A$\\
 \hline
 $F(0)$ &  0.467  & 0.145 & 0.086 & 0.447 & $-0.035$ & $-0.278$\\
 $\Lambda_1$(GeV) &  5.10  & 4.89 & 6.14 & 4.69 & 4.97 & 4.58\\
 $\Lambda_2$(GeV) &  9.03  & 5.46 & 6.28 & 12.20 & 5.05 & 7.08\\
\hline\hline
 \end{tabular}
 \end{table}

The effective Lagrangians for the $P_{cs}\rightarrow\bar{D}^*\Xi_c$ and $\bar{D}_s^{(*)}\rightarrow \bar{D}^*\bar{K}$ read
\begin{equation}\label{efflag}
\begin{split}
    \mathcal{L}_{P_{cs1}\Xi_c\bar{D}^*}&=g_{P_{cs1}\Xi_c\bar{D}^*}  \bar{\Xi}_c\gamma_5(g_{\mu\nu}-\frac{p_{\mu}p_{\nu}}{m_{Pcs}^2})\gamma^\nu
P_{cs}D^{*\mu}, \\
    \mathcal{L}_{P_{cs2}\Xi_c\bar{D}^*}&=g_{P_{cs2}\Xi_c\bar{D}^*} \bar{\Xi}_c P_{cs2\mu}D^{*\mu},\\
    \mathcal{L}_{KD_sD^*}&=ig_{KD_sD^*}D^{*\mu}[\bar{D}_s\partial_{\mu}K-(\partial_\mu \bar{D}_s)K]+H.c.,\\
    \mathcal{L}_{KD_s^*D^*}&=-g_{KD_s^*D^*} \epsilon^{\mu\nu\alpha\beta}[\partial_\mu\bar{D}_\nu^*\partial_\alpha D^*_{s\beta}\bar{K}+\partial_\mu D_\nu^*\partial_\alpha\bar{D}^*_{s\beta}K] +H.c.,
\end{split}
\end{equation}
where $P_{cs1}$ and $P_{cs2}$ denote the $P_{cs}(4459)$  state with  $J^P=\frac{1}{2}^-$ and $\frac{3}{2}^-$, respectively. The $g_{KD_sD^*}$ and $g_{KD_s^*D^*}$ are the kaon meson couplings to $D_{s}D^{\ast}$ and $D_{s}^{\ast}D^{*}$, respectively. We take $g_{KD_sD^*}=5.0$ and $g_{KD_s^*D^*}=7.0$ GeV$^{-1}$ in the present work, which are extracted from Ref.~\cite{Azevedo:2003qh}.   The $g_{P_{cs1}\Xi_c\bar{D}^*}$ and $g_{P_{cs2}\Xi_c\bar{D}^*}$ are the couplings between $P_{cs}$ and its components, whose values are not known a priori, but can be computed with the compisteness conditions~\cite{Weinberg:1962hj,Salam:1962ap,Hayashi:1967bjx} or in molecular models, e.g., Refs.~\cite{Xiao:2021rgp}, or in lattice QCD.

With the effective Lagrangians above, the decay amplitudes for $\Xi_{b}(p)\to \bar{D}_s^{(*)}(p_1)\Xi_c(p_2)[\bar{D}^*(q)]\rightarrow \bar{K}(p_3)P_{cs1}(p_4)$ read
\begin{eqnarray}\label{diabc}
\mathcal{M}_{P_{cs1}}&=&\mathcal{M}_{\bar{D}_s}^{P_{cs1}} + \mathcal{M}_{\bar{D}_{s}^*}^{P_{cs1}},\nonumber\\
\mathcal{M}_{\bar{D}_s}^{P_{cs1}}&=&i^3 \int\frac{d^4 q}{(2\pi)^4}[g_{p_{cs1}\Xi_c \bar{D}^\ast}\bar{u}(p_4)\gamma^\nu \gamma_5(g_{\mu\nu}-\frac{p_{4\mu}p_{4\nu}}{m^2_4})](p_2\!\!\!\!\!\slash+m_2)\nonumber\\
&&\times[i(A+B\gamma_5)u(p)][-g_{KD^\ast D_s}(p_1+p_3)_\alpha](-g^{\mu\alpha}+\frac{q^\mu q^\alpha}{m^2_E}) \nonumber\\
&&\times\frac{1}{p^2_1-m^2_1}\frac{1}{p^2_2-m^2_2}\frac{1}{q^2-m^2_E}\mathcal{F}(q^2,m_E^2),\nonumber\\
\mathcal{M}_{\bar{D}_{s}^*}^{P_{cs1}}&=&i^3 \int\frac{d^4 q}{(2\pi)^4}[g_{p_{cs1}\Xi_c \bar{D}^\ast}\bar{u}(p_4)\gamma^\nu \gamma_5(g_{\mu\nu}-\frac{p_{4\mu}p_{4\nu}}{m^2_4})](p_2\!\!\!\!\!\slash+m_2)\nonumber\\
&&\times[(A_1 \gamma_\alpha \gamma_5+A_2\frac{p_{2\alpha}}{m}\gamma_5+B_1 \gamma_\alpha+B_2\frac{p_{2\alpha}}{m})u(p)]\nonumber\\
&&\times[-g_{KD^\ast D^\ast_s}\varepsilon_{\rho\lambda\eta\tau}q^\rho p^\eta_1](-g^{\mu\lambda}+\frac{q^{\mu}q^{\lambda}}{m^2_E})(-g^{\alpha\tau}+\frac{p_1^{\alpha}p_1^{\tau}}{m^2_1})\nonumber\\
&&\times\frac{1}{p^2_1-m^2_1}\frac{1}{p^2_2-m^2_2}\frac{1}{q^2-m^2_E}\mathcal{F}(q^2,m_E^2).
\end{eqnarray}

The decay amplitudes of $\Xi_b(p)\rightarrow \bar{D}^{(\ast)}_s(p_1)\Xi_c(p_2)[\bar{D}^\ast(q)]\rightarrow \bar{K}(p_3)P_{cs2}(p_4)$ read
\begin{eqnarray}\label{diade}
\mathcal{M}_{P_{cs2}}&=&\mathcal{M}_{\bar{D}_s}^{P_{cs2}} + \mathcal{M}_{\bar{D}_s^*}^{P_{cs2}},\nonumber\\
\mathcal{M}_{\bar{D}_s}^{P_{cs2}}&=&i^3 \int\frac{d^4 q}{(2\pi)^4}[-ig_{p_{cs2}\Xi_c \bar{D}^\ast}\bar{u}_\mu(p_4)](p_2\!\!\!\!\!\slash+m_2)[i(A+B\gamma_5)\nonumber\\
&&u(p)][-g_{KD^\ast D_s}(p_1+p_3)_\nu](-g^{\mu\nu}+\frac{q^{\mu}q^{\nu}}{m^2_E})\nonumber\\
&&\times\frac{1}{p^2_1-m^2_1}\frac{1}{p^2_2-m^2_2}\frac{1}{q^2-m^2_E}\mathcal{F}(q^2,m_E^2), \nonumber\\
\mathcal{M}_{\bar{D}_s^*}^{P_{cs2}}&=&i^3 \int\frac{d^4 q}{(2\pi)^4}[-ig_{p_{cs2}\Xi_c \bar{D}^\ast}\bar{u}_\sigma(p_4)](p_2\!\!\!\!\!\slash+m_2)\nonumber\\
&&\times[(A_1 \gamma_\rho \gamma_5+A_2\frac{p_{2\rho}}{m}\gamma_5+B_1 \gamma_\rho+B_2\frac{p_{2\rho}}{m})u(p)]\nonumber\\
&&\times[-g_{KD^\ast D^\ast_s}\varepsilon_{\mu\nu\alpha\beta}q^\mu p^\alpha_1](-g^{\sigma\nu}+\frac{q^\sigma q^\nu}{m^2_E})(-g^{\rho\beta}+\frac{p^\rho_1 p^\beta_1}{m^2_1})\nonumber\\
&&\times\frac{1}{p^2_1-m^2_1}\frac{1}{p^2_2-m^2_2}\frac{1}{q^2-m^2_E}\mathcal{F}(q^2,m_E^2).
\end{eqnarray}

We follow  Ref.~\cite{Wu:2019rog} and introduce a monopole form factor to depict the off-shell effect of the exchanged $\bar{D}^*$ mesons,
\begin{eqnarray}
\mathcal{F}(q^2,m^2) =\frac{m^2 -\Lambda^2}{q^2-\Lambda^2},
\end{eqnarray}
where $\Lambda=m+\alpha \Lambda_{QCD}$ with $\Lambda_{QCD}=220 \ \mathrm{MeV}$, and $\alpha$ is a model parameter. In this way, the triangle diagrams are free of any ultraviolet divergence. Collecting all the pieces together, the decay width for $\Xi_b\rightarrow P_{cs}\bar{K}$ could be expressed as
\begin{eqnarray}
   \Gamma = \frac{1}{2J+1}\frac{1}{8\pi}\frac{1}{m_{\Xi_b}^2}|\vec{p}|\sum|\mathcal{M}_{P_{cs1}/P_{cs2}}|^2,
\end{eqnarray}
where  $|\vec{p}|$ denotes the momentum of $\bar{K}$ or $P_{cs}$ in the rest frame of $\Xi_b$.

\begin{figure}[htpb]
\begin{center}
\begin{tabular}{cc}
\begin{minipage}[t]{0.45\linewidth}
\begin{center}
\begin{overpic}[scale=.6]{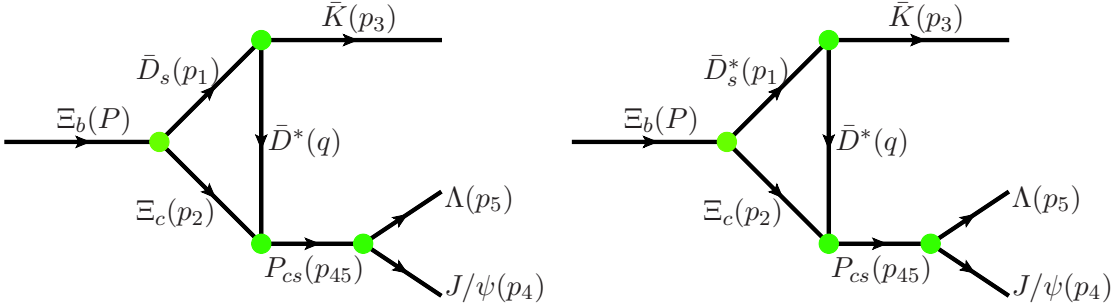}
		\put(59,5){$P_{cs}(p_{45})$}
		\put(100,21){$\Lambda(p_{5})$}
    	\put(100,1){$J/\psi(p_{4})$}
		
		\put(30,18){$\Xi_{c}(p_{2})$}
		
		\put(30,50){$\bar{D}_{s}(p_{1})$}
		
		\put(12,39){$\Xi_{b}(P)$ }
		\put(72,62){$\bar{K}(p_{3})$} \put(60,34){$\bar{D}^{\ast}(q)$}
\end{overpic}
\end{center}
\end{minipage}
&
\begin{minipage}[t]{0.45\linewidth}
\begin{center}
\begin{overpic}[scale=0.6]{threebody.eps}
		\put(59,5){$P_{cs}(p_{45})$}
		\put(100,21){$\Lambda(p_{5})$}
    	\put(100,1){$J/\psi(p_{4})$}
		
		\put(30,18){$\Xi_{c}(p_{2})$}
		
		\put(30,50){$\bar{D}_{s}^{\ast}(p_{1})$}
		
		\put(12,39){$\Xi_{b}(P)$ }
		\put(72,62){$\bar{K}(p_{3})$} \put(60,34){$\bar{D}^{\ast}(q)$}
\end{overpic}
\end{center}
\end{minipage}
\end{tabular}
\caption{Feynman diagrams for  the $\Xi^-_b \to (P_{cs}\to)J/\psi\Lambda K^-$ decay.}
\label{lineshape}
\end{center}
\end{figure}


\subsection{$J/\psi \Lambda$ invariant mass distribution of the $\Xi_b^-\to K^- P_{cs}\to K^- J/\psi\Lambda$}

With the weak decay vertices described in Eq.~(\ref{Eq:weakdecayV1}) and Eq.~(\ref{Eq:weakdecayV2}), we can further work out the invariant mass distribution of the $\Xi_b^-\rightarrow J/\psi\Lambda K^-$ decay. Parameterizing the intermediate $P_{cs}(4459)$ state with a Breit-Wigner resonance,   the amplitudes of Fig.~\ref{lineshape} read
\begin{equation}
\begin{split}
    \mathcal{A}=&\mathcal{A}_{\bar{D}_s}+\mathcal{A}_{\bar{D}_s^*},\\
    \mathcal{A}_{\bar{D}_s}=&i\int\frac{d^4 q}{(2\pi)^4}\bar{u}(p_5)\epsilon^\mu(p_4)\mathcal{A}_{\mu \alpha P_{cs1}/P_{cs2}}\cdot(\slashed{p}_2+m_2)\cdot \mathcal{A}_{\Xi_b\Xi_c\bar{D}_s} u(P) \\
    &\times\frac{ \mathcal{A}^\alpha_{KD_sD^*}\cdot(-g_{\mu\alpha}+\frac{q^\mu q^\alpha}{m_E^2})}{(p_2^2-m_2^2)(p_1^2-m_1^2)(q^2-m_E^2)}\mathcal{F}(q^2,m_E^2), \\
    \mathcal{A}_{\bar{D}_s^*}=&i\int\frac{d^4 q}{(2\pi)^4}\bar{u}(p_5)\epsilon^\mu(p_4) \mathcal{A}_{\mu \alpha P_{cs1}/P_{cs2}}\cdot(\slashed{p}_2+m_2)\cdot \mathcal{A}^\beta_{\Xi_b\Xi_c\bar{D}_s^*} u(P) \\
    &\times\frac{ \mathcal{A}^{\nu\beta}_{KD_s^*D^*}\cdot(-g_{\mu\nu}+\frac{q^\mu q^\nu}{m_E^2})(-g_{\alpha\beta}+\frac{q^\alpha q^\beta}{m_1^2})}{(p_2^2-m_2^2)(p_1^2-m_1^2)(q^2-m_E^2)}\mathcal{F}(q^2,m_E^2),
\end{split}
\end{equation}
where $(p_4+p_5)^2=p_{45}^2=M_{45}^2$ denotes the invariant mass of the $J/\psi\Lambda$ final state and
\begin{equation}
\begin{split}
    &\mathcal{A}_{\Xi_b\Xi_c\bar{D}_s}=i(A+B\gamma_5),\\
    &\mathcal{A}^\rho_{\Xi_b\Xi_c\bar{D}_s^*}=A_1\gamma_\rho\gamma_5+A_2\frac{p_{2\rho}}{m}\gamma_5+B_1\gamma_\rho+B_2\frac{p_{2\rho}}{m},\\
    &\mathcal{A}^{\alpha}_{KD_sD^*}=-g_{KD_sD^*}(p_1^\alpha+p_3^\alpha),\\
    &\mathcal{A}^{\nu\beta}_{KD_s^*D^*}=-g_{KD_s^*D^*}\epsilon_{\mu\nu\alpha\beta}(q^\mu p_1^{\alpha}),\\
    &\mathcal{A}^{\mu a}_{P_{cs1}}=\frac{g_{P_{cs1}\Xi_c\bar{D}^*}g_{P_{cs1}J/\psi\Lambda}}{p_{45}^2-m_{P_{cs}^2}+i\Gamma_{P_{cs}}m_{P{cs}}}\gamma_5(g^{\mu\nu}-\frac{p_{45}^\mu p_{45}^\nu}{m_{P_{cs}}^2})\gamma_\nu\cdot(\slashed{p}_{45}+m_{P_{cs}})\cdot \gamma_5(g_{ab}-\frac{p_{45}^a p_{45}^b}{m_{P_{cs}}^2})\gamma_b ,\\
    &\mathcal{A}^{\mu \nu}_{P_{cs2}}=\frac{g_{P_{cs2}\Xi_c\bar{D}^*}g_{P_{cs2}J/\psi\Lambda}}{p_{45}^2-m_{P_{cs}^2}+i\Gamma_{P_{cs}}m_{P_{cs}}}(\slashed{p}_{45}+m_{P_{cs}})\cdot(-g^{\mu\nu}+\frac{\gamma^\mu\gamma^\nu}{d-1}+\frac{\gamma^\mu p_{45}^\nu-\gamma^\nu p_{45}^\mu}{(d-1)m_{P_{cs}}}+\frac{d-2}{(d-1)m_{P_{cs}}^2}p_{45}^\mu p_{45}^\nu).\nonumber
\end{split}
\end{equation}

The partial decay rate for $\Xi_b\rightarrow J/\psi\Lambda \bar{K}$ as a function of  the invariant mass $M_{J/\psi\Lambda}$ then reads
\begin{equation}
   \frac{d\Gamma}{dM_{J/\psi\Lambda}}=\frac{1}{2J+1}\frac{1}{64\pi^3}\frac{1}{m_{\Xi_b}^2}|p_3^*||p_4|\sum|\mathcal{A}|^2,
\end{equation}
with
\begin{equation}
\begin{split}
   p_3^*&=\frac{\sqrt{(m_{\Xi_b}^2-(m_K-M_{J/\psi\Lambda})^2)(m_{\Xi_b}^2-(m_K+M_{J/\psi\Lambda})^2)}}{2m_{\Xi_b}}, \\
   p_4  &=\frac{\sqrt{(M_{J/\psi\Lambda}^2-(m_{J/\psi}-m_\Lambda)^2)(M_{J/\psi\Lambda}^2-(m_{J/\psi}+m_\Lambda)^2)}}{2M_{J/\psi\Lambda}}.
\end{split}
\end{equation}

\section{Results and discussions}
\label{numcal}
In this section, we explore the decay mechanism proposed in this work. We divide our discussions into two categories, those only depend on the decay mechanism explored and those depend on a particular molecular model.

\begin{figure*}[htpb]
\centering
\includegraphics[width=0.6\textwidth]{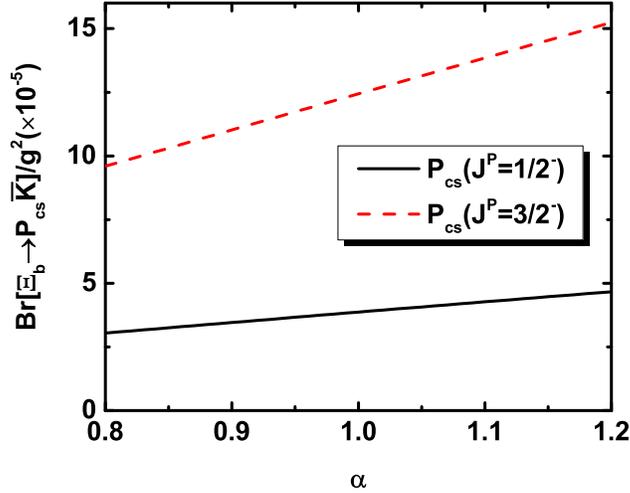}\\
\caption{Dependence of  the branching ratios $\mathrm{Br}[\Xi_b\rightarrow P_{cs}\bar{K}]$ on $\alpha$.}
\label{alphadependence}
\end{figure*}

In our framework, the parameter $\alpha$ is not known, though its value is often assumed to be about 1~\cite{Tornqvist:1993vu, Tornqvist:1993ng, Locher:1993cc, Li:1996yn}. Therefore, we first study how the calculated branching ratios depend on  the value of $\alpha$.
Varying $\alpha$ from 0.8 to 1.2, we plot the values of $\mathrm{Br}[\Xi_b\rightarrow P_{cs1}(P_{cs2})\bar{K}]/g^2_{P_{cs1/2}\Xi_c\bar{D}^*}$ in Fig.~\ref{alphadependence}. One can see that the branching ratios for $P_{cs1}$ and $P_{cs2}$ are moderately  sensitive to the value of $\alpha$ in the range studied. As a consequence, in the following, we will take $\alpha=1.0\pm0.1$ to take into account the uncertainties from $\alpha$.

\subsection{Model independent predictions}

To compute the absolute branching ratio $\mathrm{Br}[\Xi_b\to P_{cs}\bar{K}]$, we need to know the coupling constants $g_{P_{cs1}\Xi_c\bar{D}^*}$ and $g_{P_{cs2}\Xi_c\bar{D}^*}$. They can be determined model independently with the compositeness conditions~\cite{Weinberg:1962hj,Salam:1962ap,Hayashi:1967bjx}, as was done in, e.g., Ref.~\cite{Xiao:2019mvs} for the pentaquark states. With the experimental mass of $P_{cs}$, the couplings read $g_{P_{cs1}\Xi_c\bar{D}^*}=1.59$ and $g_{P_{cs2}\Xi_c\bar{D}^*}=2.76$, corresponding to a cutoff $\Lambda=1.0$ GeV (more details can be found in the Appendix). With these couplings, we find, surprisingly, that the branching ratio  for the $P_{cs}$ state with $J^P=3/2^-$ is approximately one order of magnitude larger than that for the $P_{cs}$ state with  $J^P=1/2^-$, which are
\begin{eqnarray}
\begin{split}
   \mathrm{Br}[\Xi_b\rightarrow P_{cs1} \bar{K}] &= (9.84\pm1.04) \times 10^{-5}, \\
   \mathrm{Br}[\Xi_b\rightarrow P_{cs2} \bar{K}] &= (9.48\pm1.08) \times 10^{-4}.
\end{split}\label{branchingratio}
\end{eqnarray}

In addition, using the experimental branching ratio $\mathrm{Br}[\Xi_b\rightarrow J/\psi\Lambda \bar{K}]=(2.31\pm1.37) \times 10^{-4}$ (see the Appendix how to derive this), the mass, width, and branching ratio $R$  of the $P_{cs}$ state given in Table I as inputs, we can provide a model independent constraint on the product of the two couplings in Eq.~(\ref{efflag}), $g_{P_{cs}\bar{D}^*\Xi_c}$ and $g_{P_{cs}J/\psi\Lambda}$, within the decay mechanism studied in the present work. The experimental branching ratio given in Table I is $R=2.7^{+2.0}_{-1.4}\%$. Using the formalism detailed in Sec.IIB, we obtain
\begin{eqnarray}\label{Eq:constrcoupling}
\begin{split}
g_{P_{cs}\Xi_c\bar{D}^*} g_{P_{cs}J/\psi\Lambda}= \left \{
\begin{array}{ll}
    0.18 ^{+0.10}_{-0.08} \quad\mbox{for}\quad J^P=\frac{1}{2}^-\\
    0.17 ^{+0.10}_{-0.08} \quad\mbox{for}\quad J^P=\frac{3}{2}^-
\end{array}
\right. .
\end{split}
\end{eqnarray}
The above product can be used to constrain molecular models.

\subsection{Comparison with models}

In order to produce the branching ratio $R$ defined in the introduction, in addition to the information derived above, we need to know the partial decay width of $P_{cs}$ into $J/\psi\Lambda$. For this, we turn to specific molecular models. In the following, we study the unitary approach of  Ref.~\cite{Xiao:2021rgp} and the one-boson-exchange (OBE) model of Ref.~\cite{Chen:2021tip}, calculate the branching ratio $R$, and compare with the LHCb measurement.

First, we focus on Ref.~\cite{Xiao:2021rgp}.
Note that the difference between the definition of their couplings and ours (see the Appendix for details) and  with the branching ratios $\mathrm{Br}[P_{cs}\rightarrow J/\psi\Lambda]=3.31\%$ for $P_{cs1}$ and $14.68\%$ for $P_{cs2}$ from Ref.~\cite{Xiao:2021rgp}, we obtain the couplings as $g_{P_{cs1}J/\psi\Lambda}=0.07$ and $g_{P_{cs2}J/\psi\Lambda}=0.27$. The branching ratios $R$ for the spin-parity assignment $1/2^-$ and $3/2^-$ are found to be
\begin{eqnarray}\label{eq:Rvalue}
\begin{split}
   R_{P_{cs1}}&=\frac{\mathrm{Br}[\Xi_b\rightarrow P_{cs1} \bar{K}]\mathrm{Br}[P_{cs1}\rightarrow J/\psi\Lambda]}{\mathrm{Br}[\Xi_b\rightarrow J/\psi\Lambda \bar{K}]} = 1.4\pm 0.8\%, \\
   R_{P_{cs2}}&=\frac{\mathrm{Br}[\Xi_b\rightarrow P_{cs2} \bar{K}]\mathrm{Br}[P_{cs2}\rightarrow J/\psi\Lambda]}{\mathrm{Br}[\Xi_b\rightarrow J/\psi\Lambda \bar{K}]} = 60.3\pm36.4 \% .
\end{split}
\end{eqnarray}

\begin{figure*}[h]
\centering
\includegraphics[width=0.6\textwidth]{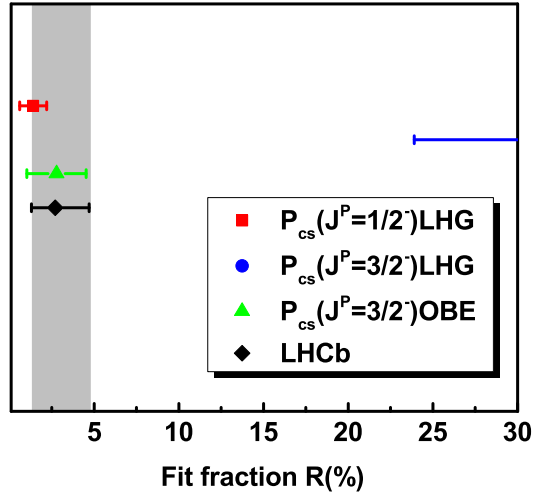}\\
\caption{Branching ratios $R$ for $P_{cs1}(J^P=1/2^-)$ and $P_{cs2}(J^P=3/2^-)$. The red square and blue circle denote our results given in Eq.~(\ref{eq:Rvalue}), while the black diamond denotes  the LHCb measurement~\cite{Aaij:2020gdg}. The results with the partial decay width obtained from the OBE model of Ref.~\cite{Chen:2021tip} (green triangle) is also shown for comparison.}
\label{ratioEXP}
\end{figure*}

In the one-boson-exchange (OBE) model of Ref.~\cite{Chen:2021tip}, the $P_{cs}(4459)$ state is interpreted as a $J^P=3/2^-$ molecular state and the partial decay width of $P_{cs}\rightarrow J/\psi\Lambda$ is estimated to be $0.06\sim0.2$ MeV. The main decay mode is found to be $P_{cs}\rightarrow K^*\Xi(\omega\Lambda)$, which accounts for $80\%$ of the total decay width.  These numbers lead to an even smaller branching ratio $\mathrm{Br}[P_{cs}\rightarrow J/\psi\Lambda]=0.6\%-0.8\%$ corresponding to the total decay width ranging from $10$ to $25$ MeV. For the coupling between the $P_{cs}$ state with $J^P=3/2^-$ and its components, we adopt the value $g_{P_{cs2}\Xi_c\bar{D}^*}=2.76$ obtained from the compositeness condition.
Using $0.7\%$ as the central value for $\mathrm{Br}[P_{cs}\rightarrow J/\psi\Lambda]$ and $0.1\%$ as its error, we obtain
\begin{equation}
     R_{P_{cs2}}=\frac{\mathrm{Br}[\Xi_b\rightarrow P_{cs2} \bar{K}]\mathrm{Br}[P_{cs2}\rightarrow J/\psi\Lambda]}{\mathrm{Br}[\Xi_b\rightarrow J/\psi\Lambda \bar{K}]} = 2.87\pm1.75 \% .
\end{equation}

All these numbers are compared with the LHCb measurement in Fig.~\ref{ratioEXP}. It is clear that the result of the OBE model seems to agree with the experimental measurement, as well as the $J^P=1/2$ case of the unitary approach. The predicted branching ratio for $J^P=3/2$ in the unitary approach is however much larger than the experimental number.

\begin{figure*}[h]
\centering
\begin{tabular}{cc}
{\includegraphics[width=0.5\textwidth]{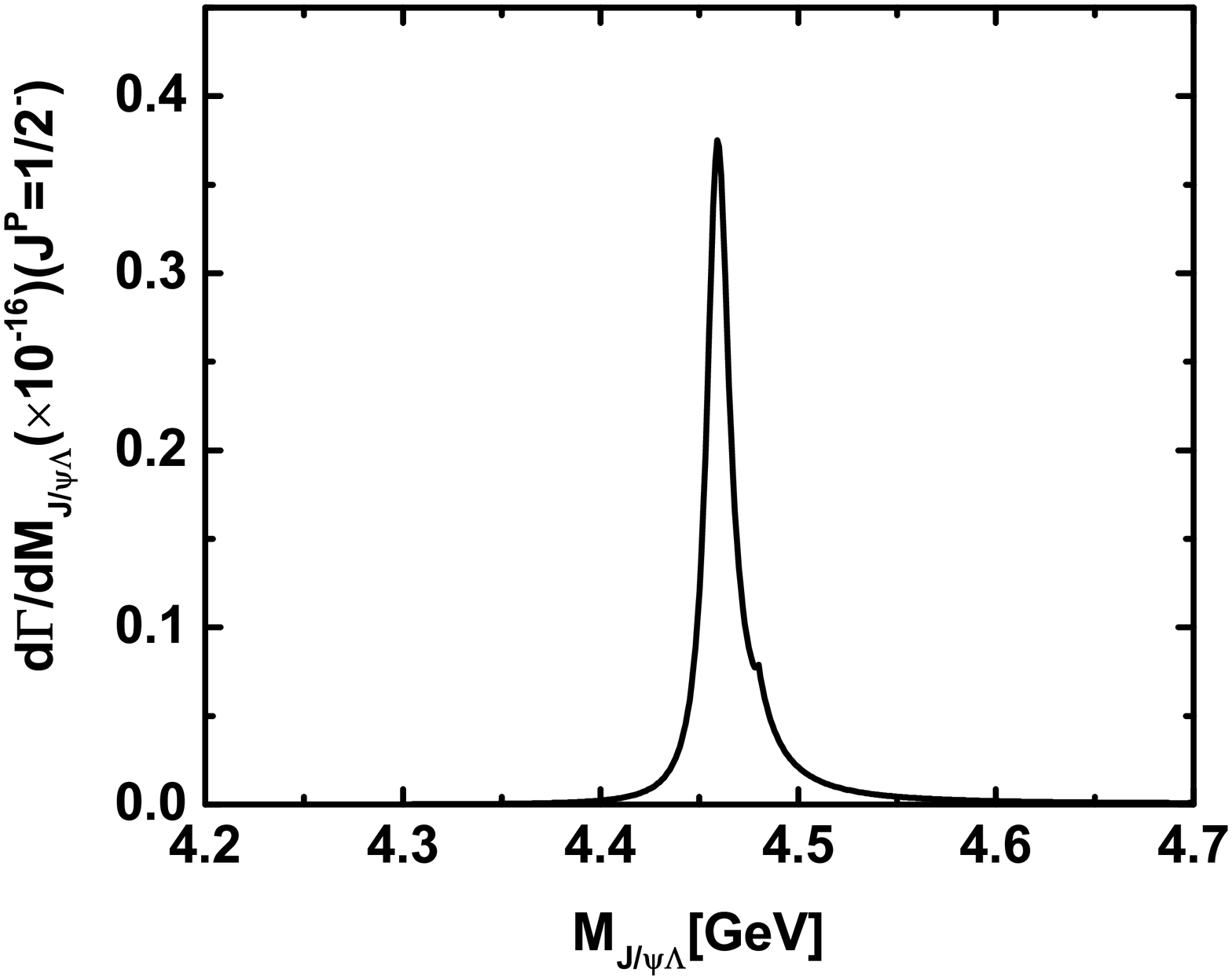}}&{\includegraphics[width=0.5\textwidth]{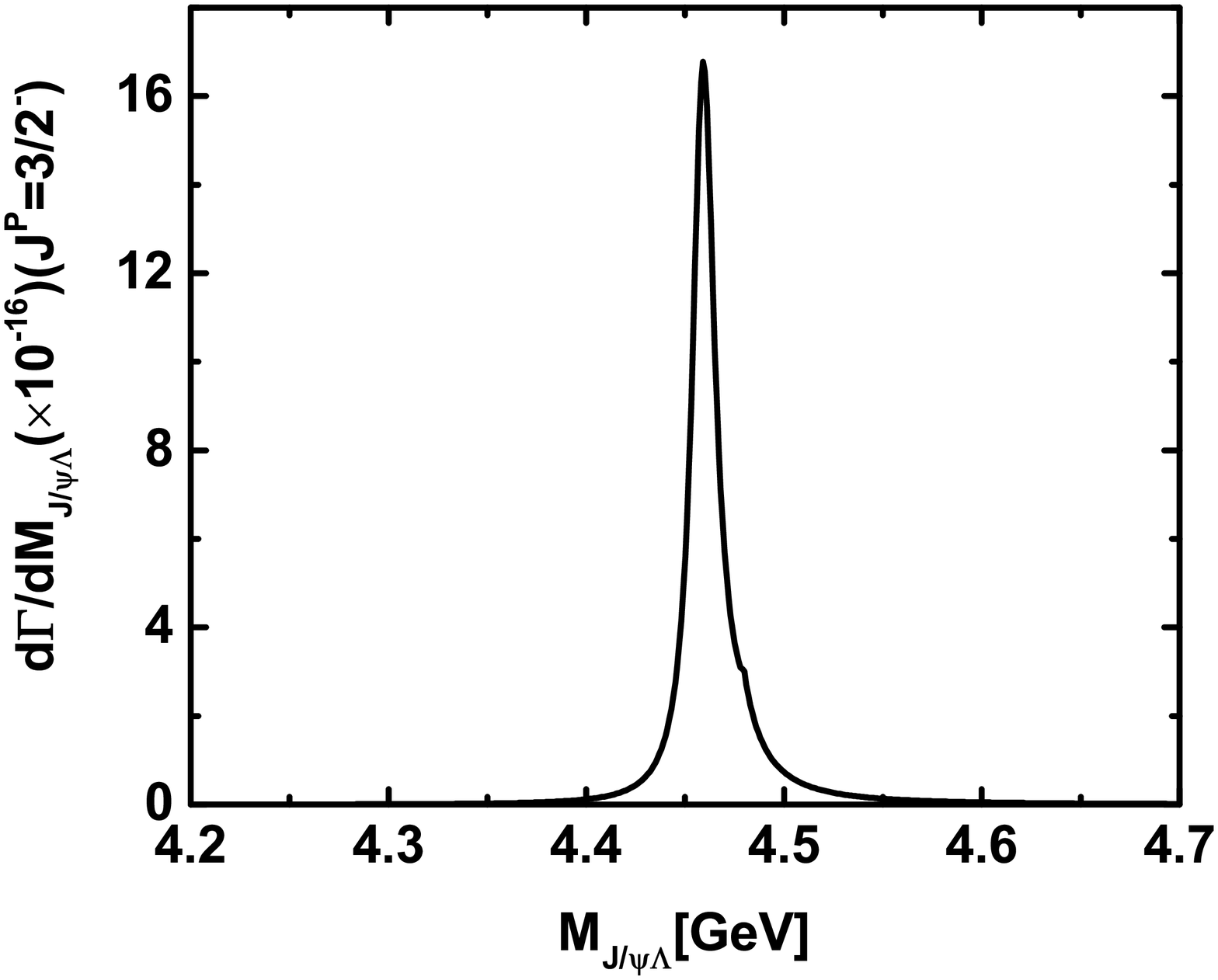}}\\
\end{tabular}
\caption{Invariant mass distribution of $\Xi_b^-\to P_{cs}K^-\to J/\psi\Lambda K^-$ for  $P_{cs}$ with $J^P=1/2^-$ (left) and $J^P=3/2^-$ (right).}
\label{differentialCS}
\end{figure*}

Finally, in Fig.~\ref{differentialCS}, we show the $J/\psi\Lambda$ invariant mass distribution of the $\Xi_b^-\to J/\psi \Lambda K^-$ decay with all the relevant couplings provided by the unitary approach of Ref.~\cite{Xiao:2021rgp} (see the Appendix for more details). They might be useful for future experimentla searches.

\section{Summary}
In this work, we studied the decay of $\Xi_b^-\to P_{cs}K^-\to J/\psi\Lambda K^-$ via a triangle mechanism.  The decay consists of three steps. First, $\Xi_b^-$ decays weakly into $\bar{D}_s^{(*)}$ and $\Xi_c$ via the external $W$-emission diagram. Using the relevant form factors determined in the relativistic quark-diquark model, this weak interaction part can be computed without any free parameters. Followed by the creation of $\bar{D}_s^{(*)}$ and $\Xi_c$ in the first step, the $\bar{D}^{(*)}_s$ state then emits a kaon and a $\bar{D}^*$. The $\bar{D}^*$ and  $\Xi_c$  interact with each other to dynamically generate the $P_{cs}(4459)$ state, which then decays into $J/\psi\Lambda$. From such a decay mechanism, we derived a constraint on the product of couplings of the $P_{cs}(4459)$ state to the $\bar{D}^*\Xi_c$ and $J/\psi\Lambda$ channels. Determining the coupling between $P_{cs}$ and the $\bar{D}^*\Xi_c$ channel using the compositeness condition, we predicted the branching ratio $\mathrm{Br}[\Xi_b^-\to P_{cs} K^-]$. These can be useful to understanding the nature of $P_{cs}(4459)$ as a molecular state.

Using the predicted couplings by the unitary approach~\cite{Xiao:2021rgp} and the one-boson exchange model of Ref.~\cite{Chen:2021tip}, we calculated the branching ratios $\mathrm{Br}[\Xi_b^-\to (P_{cs}\to)J/\psi\Lambda K^-]$. We found that in the unitary approach, the $J^P=1/2$ assignment is prefered, while the $J^P=3/2$ assignment gives a branching ratio much larger than the experimental measurement. On the other hand, the $3/2$ assignment in the one-boson-exchange model of Ref.~\cite{Chen:2021tip} yields a branching ratio in agreement with the LHCb data. This can be traced back to the drastically different partial decay width of $P_{cs}\to J/\psi\Lambda$.

In principle, the present formalism can also be utilized to study the $\Lambda_b\to J/\psi p K^-$ decay, where the four pentaquark states, $P_c(4312)$, $P_c(4380)$, $P_c(4440)$, and $P_c(4457)$, were discovered. This has been explored in Ref.~\cite{Wu:2019rog}, which, however, suffers from the fact that the weak decay $\Lambda_b \to \bar{D}_{s}^{(*)}\Sigma_c$ is suppressed because the $ud$ quark pair in $\Lambda_b$ has spin 0, but that in $\Sigma_c$ has spin 1. As a result, the relevant transition form factors are not known and therefore one could not arrive at a quantitative determination of the branching ratios. In addition, compared to the present case, the suppression of the  $\Lambda_b \to \bar{D}_{s}^{(*)}\Sigma_c$ transition indicates that other mechanisms may play a role than the external $W$-emission studied in the present work, which complicates the study a lot.

\section{Acknowledgements}
We thank Qi Wu, Dian-Yong Chen, and Li-Ming Zhang for useful communications.
This work is partly supported by the National Natural Science Foundation of China under Grants No.11735003, No.11975041,  and No.11961141004, and the fundamental Research Funds for the Central Universities.

\section{Appendix}
\subsection{Couplings from the compositeness conditions}

With the assumption that the $P_{cs}$ state observed by the LHCb Collaboration can be interpreted as a molecular state of $\bar{D}^*\Xi_c$ with $J^P=1/2^-$ or $J^P=3/2^-$, we can calculate the couplings between the $P_{cs}$ state and its components with the compositeness condition, which is quite similar to what was done in Refs.~\cite{Xiao:2019mvs,Xiao:2016hoa}.

\begin{figure}[!h]
\centering
\begin{overpic}[scale=.7]{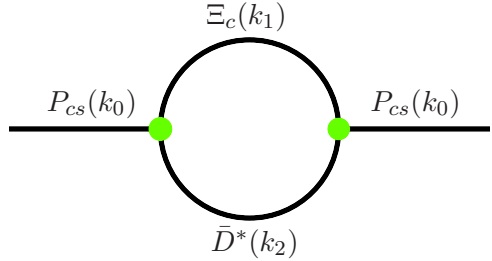}
\put(42,-6){$\bar{D}^{\ast}(k_{2})$} \put(41,41){$\Xi_{c}(k_{1})$}
\put(8,23){$P_{cs}(k_{0})$} \put(75,23){$P_{cs}(k_{0})$}
\end{overpic}
\caption{Mass operators of the $P_{c}$ }
\end{figure}

According to the compositeness rule~\cite{Weinberg:1962hj,Salam:1962ap,Hayashi:1967bjx}, the coupling constant $g_{P_{cs1/2}\Xi_c\bar{D}^*}$ can be determined from the fact that the renormalization constant of the wave function of a composite particle  should be zero. That is,
\begin{equation}
   Z_{P_{cs}}=1-\frac{d \Sigma_{P_{cs}}(k_{0})}{d
{k\!\!\!/}_0}|_{{{k\!\!\!/}_0=m_{P_{cs}}}}=0,
\end{equation}
where $\Sigma_{P_{cs}}$ denotes the self-energy of $P_{cs1}$ and $P_{cs2}$. Applying the effective Lagrangians listed in Eq.~({\ref{efflag}}), the self-energy $\Sigma_{P_{cs1/2}}$ reads
\begin{equation}
\begin{gathered}
   \Sigma_{P_{cs1}}(k_{0})=g_{P_{cs1}\Xi_c\bar{D}^*}^2 \int \frac{d^4
k_{1}}{i(2\pi)^4}\{\Phi^2[-(k_1-k_0
\omega_{\Xi_{c}})^2]\mathcal{A}_{P_{cs1}}^\mu\frac{1}{{k \!\!\!/}_{1}-m_{\Xi_{c}}}\mathcal{A}_{P_{cs1}}^\nu\frac{-g^{\mu\nu}+\frac{k_2^{\mu}k_2^{\nu}}{m_{D^{\ast}}^2}}{k_2^2-m_{D^{\ast}}^2},
\\
\Sigma_{P_{cs2}}^{\mu\nu}(k_{0})=g_{P_{cs2}\Xi_c\bar{D}^*}^2 \int \frac{d^4
k_{1}}{i(2\pi)^4}\{\Phi^2[-(k_1-k_0
\omega_{\Xi_{c}})^2]\frac{1}{{k \!\!\!/}_{1}-m_{\Xi_{c}}}\frac{-g^{\mu\nu}+\frac{k_2^{\mu}k_2^{\nu}}{m_{D^{\ast}}^2}}{k_2^2-m_{D^{\ast}}^2},
\end{gathered}
\end{equation}
with
\begin{equation}
\begin{gathered}
   \omega_{\Xi_{c}}=\frac{m_{\Xi_c}}{m_{\Xi_c}+m_{D^*}}, \\
   \mathcal{A}_{P_{cs1}}^\mu=\gamma_5(g^{\mu\nu}-\frac{k_0^\mu k_0^\nu}{m_{P_{cs}}^2})\gamma_\nu.
\end{gathered}
\end{equation}
The $\Phi[-p^2]=\mathrm{exp}(p^2/\Lambda^2)$ is the Fourier transformation of the correlation in the Gaussian form with $\Lambda$ being the size parameter which characterizes the distribution of components inside the molecule. With all the formula above and taking $\Lambda=1.0$ GeV, we obtain the couplings between the $P_{cs}$ states and $\bar{D}^*\Xi_c$, which are $g_{P_{cs1}\Xi_c\bar{D}^*}=1.59$ for $J^P=1/2^-$ and $g_{P_{cs2}\Xi_c\bar{D}^*}=2.76$ for $J^P=3/2^-$.

\subsection{Determination of the branching ratio  $\mathrm{Br}[\Xi_b\rightarrow J/\psi\Lambda \bar{K}]$ }
Experimentally, the branching ratio of $\Xi_b\rightarrow J/\psi\Lambda \bar{K}$ has been measured to be~\cite{Aaij:2017bef}
\begin{eqnarray}
   \frac{f_{\Xi_b}}{f_{\Lambda_b}}\times \frac{\mathrm{Br}[\Xi_b\rightarrow J/\psi \Lambda \bar{K}]}{\mathrm{Br}[\Lambda_b\rightarrow J/\psi\Lambda]}= (4.19\pm0.29\pm0.15) \times 10^{-2},
\end{eqnarray}
where $f_{\Xi_b}$ and $f_{\Lambda_b}$ refer to the $b$ quark fragmentation fractions  into $\Xi_b^-$ and $\Lambda_b^0$, the ratio of which is~\cite{Aaij:2019ezy}
\begin{eqnarray}
   \frac{f_{\Xi_b}}{f_{\Lambda_b}}= (6.7\pm0.5\pm0.5\pm2.0) \times 10^{-2} ,
\end{eqnarray}
while the branching ratio of $\Lambda_b \rightarrow J/\psi \Lambda$ has been measured by the CDF Collaboration~\cite{Abe:1996tr}
\begin{eqnarray}
   \mathrm{Br}[\Lambda_b\rightarrow J/\psi\Lambda]= (3.7\pm1.7\pm0.7) \times 10^{-4} .
\end{eqnarray}
With all the ratios given above, one can compute the branching ratio of $\Xi_b\rightarrow J/\psi\Lambda \bar{K}$
\begin{eqnarray}
   \mathrm{Br}[\Xi_b\rightarrow J/\psi\Lambda \bar{K}]= (2.31\pm1.37) \times 10^{-4} .
\end{eqnarray}
The large uncertainty can be traced back to the experimental uncertainty in the braching ratio $\mathrm{Br}[\Lambda_b\rightarrow J/\psi\Lambda]$, which accounts for about $50\%$, and the large uncertainty in the ratio of fragmentation fractions coming from the estimation of SU(3) breaking effects~\cite{Aaij:2019ezy}.

\subsection{Couplings from the unitary approach}
In our convention,  the $J/\psi\Lambda$ partial decay widths of  the $P_{cs}$ state with $J^P=1/2^-$ and $3/2^-$ are expressed as
\begin{equation}
\begin{gathered}
   \Gamma_{P_{cs1}\rightarrow J/\psi\Lambda}=\frac{1}{2}\frac{g_{P_{cs1}J/\psi\Lambda}^2}{8\pi}\frac{1}{m_{P_{cs}}^2}|q|\sum |\mathcal{A}_{P_{cs1}}|^2, \\
   \Gamma_{P_{cs2}\rightarrow J/\psi\Lambda}=\frac{1}{4}\frac{g_{P_{cs2}J/\psi\Lambda}^2}{8\pi}\frac{1}{m_{P_{cs}}^2}|q|\sum |\mathcal{A}_{P_{cs2}}|^2,
\end{gathered}
\end{equation}
where the modules of amplitude squared are
\begin{equation}
\begin{gathered}
   \sum |\mathcal{A}_{P_{cs1}}|^2=\frac{\left((m_{\Lambda }+m_{P_{\text{cs}}})^2-m_{J/\psi }^2\right) \left(\left(m_{J/\psi }^2-m_{\Lambda }^2\right)^2+2 m_{P_{\text{cs}}}^2 \left(5 m_{J/\psi
   }^2-m_{\Lambda }^2\right)+m_{P_{\text{cs}}}^4\right)}{2 m_{J/\psi }^2  m_{P_{\text{cs}}}^2}, \\
   \sum |\mathcal{A}_{P_{cs2}}|^2=\frac{\left((m_{\Lambda }+m_{P_{\text{cs}}})^2-m_{J/\psi }^2\right) \left(\left(m_{J/\psi }^2-m_{\Lambda }^2\right)^2+2 m_{P_{\text{cs}}}^2 \left(5 m_{J/\psi
   }^2-m_{\Lambda }^2\right)+m_{P_{\text{cs}}}^4\right)}{3 m_{J/\psi }^2  m_{P_{\text{cs}}}^2},
\end{gathered}
\end{equation}
in which $q$ denotes the momentum of $J/\psi$ in the rest frame of the $P_{cs}$ state. Using the partial decay widths from Ref.~\cite{Xiao:2021rgp}, we obtain $g_{P_{cs1}J/\psi\Lambda}=0.07$ and $g_{P_{cs2}J/\psi\Lambda}=0.27$. Similarly, we obtain  $g_{P_{cs1}\Xi_c\bar{D}^*}=1.25$ and  $g_{P_{cs2}\Xi_c\bar{D}^*}=2.17$.

\bibliography{pentaquarks}
\end{document}